\documentclass[aps,prb,twocolumn,showpacs,superscriptaddress,nofootinbib]{revtex4-1}  
\usepackage{graphicx}  
\usepackage{dcolumn}   
\usepackage{bm}        
\usepackage{amsmath, amsthm, amssymb}
\usepackage{amsbsy} 
\usepackage{bbold} 
\usepackage{mathrsfs}
\usepackage{array}
\usepackage[usenames]{color}
\usepackage{soul} 
\usepackage{ulem} 
\usepackage{units} 
\newcommand{\ave}[1]{\langle #1\rangle}
\newcommand{\dn}{\downarrow}
\newcommand{\iv}{\mathbf i}
\newcommand{\jv}{\mathbf j}
\newcommand{\kv}{\mathbf k}
\newcommand{\lvec}[1]{\mathbf{#1}}
\newcommand{\qv}{\mathbf q}
\newcommand{\up}{\uparrow}
\newcommand{\xvu}{\mathbf{\hat{x}}}

\newcommand{\ellv}{\ensuremath{\boldsymbol\ell}}

\begin{document}

\normalem

\title{The 2D attractive Hubbard model and the BCS-BEC crossover}
\author{Rodrigo A. Fontenele, Natanael C. Costa, Raimundo R. \surname{dos Santos}, and Thereza Paiva}
\affiliation{Instituto de F\'isica, Universidade Federal do Rio de Janeiro Cx.P. 68.528, 21941-972 Rio de Janeiro RJ, Brazil}
\begin{abstract}
Recent experiments with ultracold fermionic atoms in optical lattices have provided a tuneable and clean realization of the attractive Hubbard model (AHM). 
In view of this, several physical properties may be thoroughly studied across the crossover between weak (Bardeen-Cooper-Schrieffer, BCS) and strong (Bose-Einstein condensation, BEC) couplings.
Here we report on extensive determinant Quantum Monte Carlo (DQMC) studies of the AHM on a square lattice, from which several different quantities have been calculated and should be useful as a roadmap to experiments. 
We have obtained a detailed phase diagram for the critical superconducting temperature, $T_c$, in terms of the band filling, $\ave{n}$, and interaction strength, $U$, from which we pinpoint a somewhat wide region $|U|/t \approx 5 \pm 1$ ($t$ is the hopping amplitude) and $\ave{n} \approx 0.79 \pm 0.09$ leading to a maximum $T_c \approx 0.16 t$.
Two additional temperature scales, namely pairing, $T_p$, and degeneracy, $T_d$, have been highlighted: the former sets the scale for pair formation (believed to be closely related to the scale for the gap of spin excitations in cuprates), while the latter sets the scale for dominant quantum effects.
Our DQMC data for the distribution of doubly occupied sites, for the momentum distribution function, and for the quasiparticle weight show distinctive features on both sides of the BCS-BEC crossover, being also suggestive of an underlying crossover between Fermi- and non-Fermi liquid behaviors.
\end{abstract}


\pacs{
71.10.Fd, 
74.25.Dw 
71.10.Ay  
02.70.Uu  
71.27.+a   
67.10.Db  
74.78.-w  
67.25.D-   
}
\maketitle

\section{Introduction}
\label{sec:Intro}

In its simplest form, the attractive Hubbard model (AHM) \cite{Micnas90} comprises of fermions moving in a single band (nearest neighbor hopping integral $t$) subject to an on-site interaction, $U<0$, which favors the formation of local pairs.  
Over the years this model has played an important role in describing many aspects of superconductivity. 
For instance, this model naturally contemplates Cooper pair formation within a certain temperature scale, $T_p\gtrsim T_c$, where $T_c$ is the  critical temperature for superconductivity, when pairs actually condense \cite{Randeria92,dosSantos94,Paiva10}.
This behavior, absent in the Bardeen-Cooper-Schrieffer (BCS) pairing theory \cite{Bardeen57}, has been suggested to be relevant to pseudogap phenomena in high-temperature cuprate superconductors~\cite{Wilson01}.
Another important feature of the AHM is the possibility of smoothly interpolating between two limits: from weak coupling, where one has BCS behavior, with large pair coherence length, to strong coupling, where pairs are tightly-bound, with short coherence length, and undergo Bose-Einstein condensation \cite{Micnas90,Randeria95a,Chen05,Randeria14}.

With the continuing development of optical lattices experiments, in which ultracold fermionic atoms are loaded and the interaction amongst them is controlled through an external magnetic field \cite{Jaksch05,Bloch08,Esslinger10,McKay11}, the Hubbard model has been experimentally studied in an unprecedented way; we note that here we will refer to superconductivity of \emph{neutral} ultracold atoms as their \emph{superfluidity.}
This was followed by yet another important advance, the quantum-gas microscope \cite{Bakr09}, which paved the way to visualise the atomic distribution on the lattice and draw quantitative conclusions. 
Indeed, several properties of the attractive Hubbard model on a square optical lattice were measured this way, including correlation functions \cite{Mitra18, kohl-1,kohl-2}.

Notwithstanding the progress achieved so far, several issues still need attention, both theoretically and experimentally.
First, accurate theoretical estimates for the critical temperature on the square lattice, $T_c(n,U)$, are only available for limited sets of either band filling, $n$, or $U$. 
Indeed, for $U=-4t$, data for $T_c(n, -4t)$ obtained from determinant quantum Monte Carlo (DQMC) simulations yield a maximum $T_c\approx 0.15\, t/k_B$ ($k_B$ is the Boltzmann constant, which from here on will be omitted) around $n\approx 0.7$ \cite{Paiva04}; subsequent DQMC simulations at $n= 0.7$~\cite{Paiva10}, found a maximum $T_c\approx 0.17 t$ near $U=-5t$. 
These estimates should be compared with the lowest temperatures reached so far in experiments on the AHM, namely $T \approx 0.4 t \approx 22 \,\text{nK}$ \cite{Mitra18}.
Thus, the search for an AHM `sweet spot' (i.e.\ a range of combinations of $n$ and $U$ giving rise to the maximum $T_c$) is of crucial importance to guide experimental studies of the phase transition on a square optical lattice.  

Another aspect demanding a more quantitative description is that of temperature scales such as the degeneracy temperature and the pairing temperature. 
The former sets the scale below which quantum effects dominate, while the latter is usually associated with pair formation and gap opening in spin excitations \cite{Randeria92,dosSantos94,Magierski09,Magierski11,Wlazlowski13,Tajima14}. 
Placing these temperature scales in a $T_c\times U$ phase diagram should therefore provide interesting insights.

A third point needing attention concerns the BCS-BEC crossover. 
So far, most of the experimental studies of this crossover in ultracold atoms have been carried out in the continuum \cite{Randeria14}. 
On an optical lattice one has at our disposal very accurate imaging techniques which can provide quantitative measures of double occupancy, so that a distribution of double occupancy should be very helpful to gain further quantitative insight into this crossover.
In actual fact one may envisage yet another crossover: while at weak coupling the normal phase may be described by a Fermi liquid, one should not expect such a simple behavior at strong coupling, since one has tightly bound pairs whose interactions may be thought of as being mediated by the unpaired fermions, which may be indicative of a non-Fermi liquid regime.

With the purpose of providing some quantitative insights into these unresolved issues, here we report on results of extensive determinant quantum Monte Carlo (DQMC) simulations on the attractive Hubbard model. 
The layout of the paper is as follows.
In Sec.\ \ref{sec:HQMC} we discuss the model and highlight the main aspects of DQMC, including the different quantities used to probe the physical properties of the system. 
In Sec.\ \ref{sec:results} we present the phase diagrams, which include the critical, degeneracy and pairing temperatures.
Proposals to probe the BCS-BEC crossover are discussed in Sec.\ \ref{sec:Xover}, and Sec.\ \ref{sec:conc} presents our final conclusions.
 

\section{Model and Methodology}
\label{sec:HQMC}
The attractive Hubbard Hamiltonian reads
\begin{align}\label{Eq:Hamil}
\nonumber
\mathcal{H} = &
-t\sum_{\langle \mathbf{i}, \mathbf{j} \rangle,\,\sigma} (c^{\dagger}_{\mathbf{i},\sigma} c_{\mathbf{j},\sigma} + \mathrm{H.c.})
-\mu\sum_{\mathbf{i}, \sigma} n_{\mathbf{i}, \sigma}
\\
& -|U|\sum_{\mathbf{i}} (n_{\mathbf{i}\uparrow} -\nicefrac{1}{2}) (n_{\mathbf{i}\downarrow} 
-\nicefrac{1}{2}),
\end{align}
where the sums run over sites of a square lattice, with $\langle \mathbf{i}, \mathbf{j} \rangle$ denoting nearest-neighbor sites.
$c^{\dagger}_{\mathbf{i} \sigma}$ ($c^{\phantom{\dagger}}_{\mathbf{i} \sigma}$) is a creation (annihilation) operator of an electron on a given site $\mathbf{i}$ with spin $\sigma$, and $n_{\mathbf{i}\sigma} \equiv c^{\dagger}_{\mathbf{i} \sigma} c_{\mathbf{i} \sigma}^{\phantom{\dagger}}$ being fermionic number operator in the conventional second quantization formalism.
The first term on the right hand side of Eq.\,\eqref{Eq:Hamil} describes particle hopping, with H.c.\,denoting  hermitian conjugate, while the second term controls the band filling through the chemical potential, $\mu$.
The last term corresponds to the local attractive interaction between electrons, with coupling strength $|U|$.
Here, the hopping integral $t$ sets the energy scale.

We investigate the finite temperature properties of the AHM  
by performing DQMC simulations \cite{Blankenbecler81,Hirsch83,Hirsch85,White89,dosSantos03b}.
The DQMC method is an unbiased numerical approach based on an auxiliary-field decomposition of the interaction, which maps onto a quadratic form of free fermions coupled to bosonic degrees of freedom $\mathcal{S}(\mathbf{i},\tau)$ in both spatial and (imaginary) time coordinates.
This method is based on a separation of the
non-commuting parts of the Hamiltonian by means of the Trotter-Suzuki decomposition, i.e.~
\begin{align}
	\mathcal{Z} &= \mathrm{Tr}\,e^{-\beta\widehat{\mathcal{H}}}
	= \mathrm{Tr}\,[(e^{-\Delta\tau(\widehat{\mathcal{H}}_{0} + \widehat{\mathcal{H}}_{\rm
U})})^{M}]\nonumber\\
&\thickapprox \mathrm{Tr}\,[e^{-\Delta\tau\widehat{\mathcal{H}}_{0}}e^{-\Delta\tau\widehat{\mathcal{H}}_{\rm
U}}e^{-\Delta\tau\widehat{\mathcal{H}}_{0}}e^{-\Delta\tau\widehat{\mathcal{H}}_{\rm
U}}\cdots],
\end{align}
where $\widehat{\mathcal{H}}_{0}$ contains the terms quadratic in fermion creation and 
destruction operators, while $\widehat{\mathcal{H}}_{\rm U}$ contains the quartic terms. 
We take $\beta=M \Delta\tau$, with $\Delta\tau$ being the grid of the imaginary-time coordinate axis.  
This decomposition leads to
an error proportional to $(\Delta\tau)^{2}$, which can be 
systematically reduced as $\Delta\tau
\to 0$.
Here, we choose $\Delta\tau \leq 0.1$ (depending on the temperature), which is small enough so
that systematic errors are comparable to the statistical ones (from the Monte Carlo sampling).

We collect DQMC data for several quantities probing superconductivity.
The $s$-wave pair correlation function is defined as
\begin{equation}
C_{\iv\jv}^\Delta \equiv 
	\langle b_\iv^\dagger b_\jv^{\phantom{\dagger}} 
	+ \text{H.c.}\rangle,
\label{eq:Cbibj}
\end{equation}
where
\begin{equation}
	b_\iv^{\phantom{\dagger}} \equiv  c_{\iv\downarrow}^{\phantom{\dagger}}c_{\iv\uparrow}^{\phantom{\dagger}}
	\quad\text{and}\quad 
	b_\iv^\dagger \equiv c_{\iv\uparrow}^\dagger c_{\iv\downarrow}^\dagger
\end{equation}
respectively annihilates and creates a pair at site $\iv$.
The decay of $C_{\iv\jv}^\Delta$ with the distance $r_{\iv\jv}\equiv |{\iv}-{\jv}|$ probes the resilience of pair coherence at a given temperature.
The Fourier transform of $C_{\iv\jv}^\Delta$ at $\mathbf{q}=0$ defines the \textit{s}-wave pair-field structure factor,
\begin{equation}
	P_s= 
	\langle \Delta^\dagger \Delta + \Delta \Delta^\dagger \rangle,
\label{eq:Ps-def}
\end{equation}
with
\begin{equation}
	\Delta^\dagger = \frac{1}{\sqrt {N}} \sum_\iv
	b_\iv^{^\dagger}
\label{eq:Delta}
\end{equation}
being the pair-field operator.

			
The finite-size scaling (FSS) behavior of $P_s$ is therefore obtained upon integration of $C_{\iv\jv}^\Delta$ over a two-dimensional system of linear dimension $L$ \cite{Moreo91,Paiva04},
\begin{equation}
P_s=L^{2 - \eta(T_c)} f(L/ \xi), \ \ \ L\gg 1,\ T\to T_c^+, 
\label{eq:Ps}
\end{equation}
where 
$\eta(T_c)=1/4$ \cite{Kosterlitz73,Berche02}, and
\begin{equation}
\xi \sim \exp \left[ \frac{A}{(T-T_c)^{1/2}} \right],
\label{eq:xi}
\end{equation}
with $A$ being a constant independent of temperature.

As discussed previously \cite{Paiva04}, estimates for the critical temperature obtained through an exponential correlation length, Eq.\,\eqref{eq:xi}, must be supplemented by an analysis of the helicity modulus (HM) for accuracy. 
The latter is a measure of the response of the system in the ordered phase to a `twist' of the order parameter \cite{Fisher73}, and can be expressed in terms of the current-current correlation functions as  \cite{Scalapino92,Scalapino93},
\begin{equation}
\rho_s=\frac{D_s}{4\pi e^2}=
\frac{1}{4} [\Lambda^L - \Lambda^T],
\label{Ds}
\end{equation}
where $D_s$ is the superfluid weight, and
\begin{equation}
\Lambda^L \equiv \lim_{q_x \to 0} \Lambda _{xx} (q_x, q_y=0,\omega_n=0),
\end{equation}
and
\begin{equation}
\Lambda^T \equiv \lim_{q_y \to 0} \Lambda _{xx} (q_x=0, q_y,\omega_n=0),
\end{equation}
are, respectively, the limiting longitudinal and transverse responses, 
with
\begin{equation}
	\Lambda_{xx}(\qv, \omega_n)=
 		\sum_{\ellv} \int_0^\beta d\tau\, 
		e^{i\qv \cdot \ellv} e^{i \omega_n \tau} \Lambda_{xx}(\ellv,\tau),
\label{lambdaq}
\end{equation}
where $\omega_n=2 n \pi T$; 
\begin{equation}
\label{lambda}
\Lambda_{xx}(\ellv, \tau)= \langle j_x(\ellv, \tau) j_x (0,0) \rangle,
\end{equation}
where
\begin{equation}
	j_x(\ellv,\tau)= e^{ {\cal H} \tau} 
		\left[ it\sum_\sigma 
		\left( c_{\ellv + \xvu,\sigma}^{^{\dagger}}c_{\ellv,\sigma} 
		-c_{\ellv,\sigma}^{^{\dagger}} c_{\ellv+ \xvu,\sigma}\right)
		 \right] e ^{-{\cal H} \tau}
\end{equation}
is the $x$-component of the current density operator; see Ref.\,\cite{Scalapino92} for details.

\begin{figure}[t]
\centering
\includegraphics[scale=0.28]{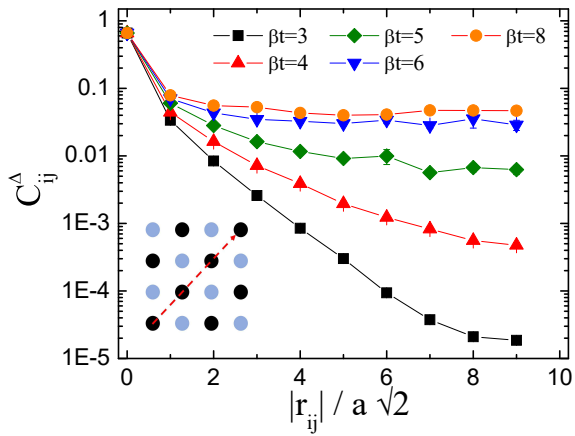} 
\caption{(Color online) Pairing correlation function as a function of distance along the diagonal direction (see inset) on an $18\times18$ lattice, for different inverse temperatures, $\beta$, with $U/t=-5$ and electronic density $\langle n\rangle=0.5$. PBC limits the farthest distance to $La/\sqrt{2}$, where $a$ is the lattice spacing.
}
\label{fig:Cij} 
\end{figure}

At the KT transition, the following universal-jump relation involving the 
helicity modulus holds \cite{Nelson77}:
\begin{equation}\label{eq:helicity_mod}
T_c = \frac {\pi} {2} \rho_s^-,
\end{equation}
where $\rho_s^-$ is the value of the helicity modulus just below the critical temperature. 
We therefore calculate both $\Lambda^L$ \emph{and} $\Lambda^T$ by DQMC simulations to obtain $\rho_s$ through Eq.\,\eqref{Ds}.
$T_c$ is then determined by plotting $\rho_s(T)$, and looking for the intercept with $2T/\pi$ \cite{Denteneer91,Denteneer93,Denteneer94,Paiva04}; see below.

For our purposes here, the magnetic properties are probed by the uniform susceptibility, 
\begin{equation}
	\chi_s=\frac{1}{N_s}\sum_{\lvec{i}\lvec{j}}
\int_0^\beta d\tau\ \ave{\mathbf{S}_{\lvec{i}}(\tau)\cdot\mathbf{S}_{\lvec{j}}(0)}.
\label{eq:magsusc}
\end{equation}
where $\mathbf{S}_{\lvec{i}}\equiv (1/2)\mathbf{m}_{\lvec{i}}$,
with the components of the magnetization operator being
\begin{subequations}
\label{eq:mags}
	\begin{eqnarray}
		m_{\lvec{i}}^x&\equiv&
		c_{\lvec{i}\up}^\dagger c_{\lvec{i}\dn}^{\phantom{\dagger}}+ c_{\lvec{i}\dn}^\dagger c_{\lvec{i}\up}^{\phantom{\dagger}},
	\label{eq:mx}\\
		m_{\lvec{i}}^y&\equiv&
		-i\left(c_{\lvec{i}\up}^\dagger c_{\lvec{i}\dn}^{\phantom{\dagger}}- c_{\lvec{i}\dn}^\dagger c_{\lvec{i}\up}^{\phantom{\dagger}}\right),
	\label{eq:my}\\
	m_{\lvec{i}}^z&\equiv&n_{\lvec{i}\up}-n_{\lvec{i}\dn},
	\label{eq:mz}
	\end{eqnarray}
\end{subequations}

Throughout this work our simulations were carried out on $L\times L$ square lattices with periodic boundary conditions (PBC), such that $L\leq 18$. 
Typically our data have been obtained after $5$-$10\times 10^{3}$ warming-up steps followed by $2-6\times 10^{5}$ sweeps for measurements, depending on the temperature, interaction strength and electronic density.

\section{Results}
\label{sec:results}

\subsection{Critical temperature}
\label{ssec:Tc}

In discussing the critical temperature, we first recall that charge-density wave (CDW) and singlet superconducting (SS) correlations are degenerate at half filling, thus leading to a three-component order parameter: by virtue of the Mermin-Wagner theorem, there is no long-range order at finite temperatures, and $T_c=0$ for any $U$.
As one dopes away from half filling, CDW correlations are suppressed but the two-component SS correlations remain, so that a Kosterlitz-Thouless (KT) transition at finite temperatures, $T_c$, takes place.

\begin{figure}[t]
\includegraphics[scale=0.38]{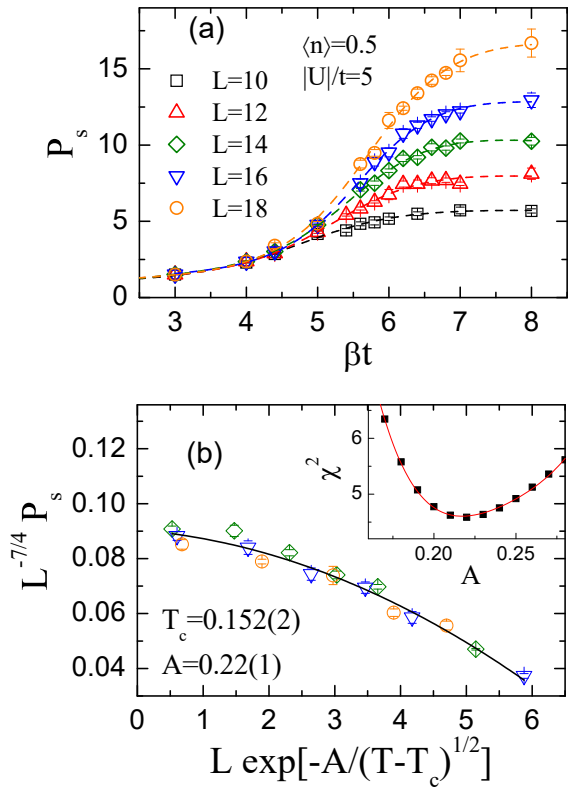} 
\caption{(Color online) (a) Pair structure factor as a function of the inverse of temperature $\beta$, for different lattice sizes, and fixed $\langle n \rangle =0.5$ and $U/t=-5$. 
(b) The data collapse of $P_{s}$ according to the Kosterlitz-Thouless FSS analysis. Inset: the chi-squared values of a polynomial fit to the data collapse, for fixed $T=0.152$ (in units of $t$).
The curves are guides to the eye.
}
\label{fig:Ps} 
\end{figure}

Let us then consider the behavior of the pairing correlation functions away from half filling, as the temperature is varied.
Figure \ref{fig:Cij} presents $C_{\iv\jv}^\Delta$ along the diagonal direction of the lattice, for fixed $\langle n \rangle=1/2$, and $U/t=-5$.
At high temperatures the steady decay of $C_{\iv\jv}^\Delta$ reflects the lack of pair coherence along the lattice, as expected. 
The situation changes completely at low temperatures, $\beta t \gtrsim 5$, with the correlations now reaching a finite value at large distances, compatible with long-range order in the ground state.
This long-range behavior is also manifested in the pairing structure factor, Eq.\,\eqref{eq:Ps-def}, as displayed in Fig.\,\ref{fig:Ps}\,(a): $P_{s}$ stabilizes at low temperatures as a result of the range of correlations being limited by the finite size of the lattice, but nonetheless experiencing a steady increase with $L$.

As mentioned in Sec.\,\ref{sec:HQMC}, we may use the FSS ansatz for $P_s$ at finite temperatures, Eq.\,\eqref{eq:Ps}, to determine $T_c$. 
Figure \ref{fig:Ps}\,(b) shows the collapse of the data appearing in panel (a), in which $T_{c}$ and $A$ are considered as independent variables, adjusted through a least squares fit. 
The inset of Fig.\,\ref{fig:Ps}\,(b) illustrates this process,
from which, by minimizing the $\chi^2$ function for a polynomial fit of the data collapse, we are able to find the most appropriate value for \textit{A}, while keeping $T_{c}$ fixed.
When this procedure is performed recursively for $T_{c}$ and $A$, we obtain the best data collapse.

\begin{figure}[t]
\includegraphics[scale=0.28]{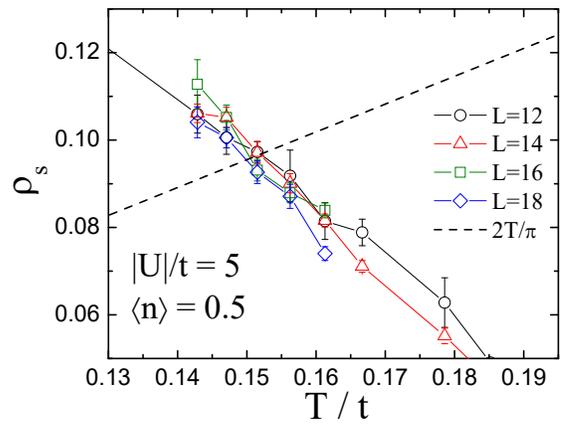} 
\caption{(Color online) Temperature behavior of the helicity modulus $\rho_{s}$ at fixed $\langle n \rangle =0.5$ and $U/t=-5$, and for several lattice sizes.}
\label{fig:rhosn05U5} 
\end{figure}

The helicity modulus provides an alternative way to estimate the critical temperature, using Eq.\,\eqref{eq:helicity_mod}, as illustrated in Fig.\,\ref{fig:rhosn05U5} for the same filling and $U$ as in Fig.\,\ref{fig:Ps}: the intersection of $\rho_s$ for each lattice size with the straight line $2T/\pi$ yields estimates for $T_c$. 
We note that the position of the intersections are not too sensitive to $L$ -- whether we take into account the scatter of all intersections shown, or just the data for the largest lattice size, the final estimate will hardly differ from $T_{c}=0.150\pm0.003$ (in units of $t$), which is in agreement with the value obtained from the data collapse.

\begin{table*}[!htb]
    \centering
    $\begin{array}{|c||c|c|c|c|c|c|}
    \hline
   \text{scale} & \ave{n} &  U/t=-3 &  U/t=-4 &  U/t=-5 &  U/t=-6 & U/t=-8 \\
     \hline
    \hline
        &0.20 & 0.061\pm 0.002 & 0.077\pm 0.006 & 0.091\pm 0.003 & 0.089\pm 0.003 &  0.077\pm 0.006\\
         &0.35 & 0.067\pm 0.004 & 0.106\pm 0.007 & 0.130\pm 0.005 & 0.128\pm 0.003 & 0.11\pm 0.01 \\
        T_c &0.50 & 0.079\pm 0.001 & 0.133\pm 0.002 & 0.152\pm 0.002 &  0.153\pm 0.002 & 0.130\pm 0.002\\
         &0.70 & 0.105\pm 0.006 & 0.149\pm 0.002 & 0.164\pm 0.003 & 0.161\pm 0.005 & 0.139\pm 0.004 \\
         &0.87 & 0.114\pm 0.003 & 0.152\pm 0.002 & 0.164\pm 0.003 & 0.157\pm 0.001 & 0.137\pm 0.002\\
           \hline\hline
        &0.35 & 0.400\pm 0.050 & 0.725\pm 0.059 & 1.338\pm 0.089 & 1.622\pm 0.119 & 2.601\pm 0.138 \\
 T_p &0.50 & 0.543\pm 0.028 & 0.725\pm 0.027 & 1.253\pm 0.089 &  1.731\pm 0.208 & 2.667\pm 0.138 \\
        &0.70 & 0.435\pm 0.050 & 0.842\pm 0.083 &   1.228 \pm 0.125 & 1.892\pm 0.074 & 2.789\pm 0.312 \\
        &0.87 & 0.498\pm 0.050 & 0.764\pm 0.027 & 1.433\pm 0.286 & 1.993\pm 0.167 & 2.791\pm 0.470\\
        \hline\hline
        &0.35 & 1.180\pm 0.069 & 1.055\pm 0.055 & 0.954\pm 0.045 & 0.871\pm 0.037 & 0.611\pm 0.055 \\
        T_d&0.50 & 1.548 \pm 0.119 & 1.547 \pm 0.119 & 1.339\pm 0.089 &  1.180 \pm 0.069 & 1.055 \pm 0.055 \\
      &  0.70 & 2.386 \pm 0.113 & 2.223 \pm 0.049 & 2.129 \pm 0.045 & 2.128 \pm 0.045 & 1.961 \pm 0.038 \\
        &0.87 & 3.033 \pm 0.092 & 3.033 \pm 0.092 & 3.033 \pm 0.092 & 3.033 \pm 0.092 & 3.033 \pm 0.092 \\
        \hline
    \end{array}$
    \caption{Superconducting critical temperature, $T_c$, pairing temperature, $T_p$, and degeneracy temperature, $T_d$, (all in units of $t$) for different fermionic densities (rows) and different strengths of attraction (columns).}
    \label{tab:allT}
\end{table*}

\begin{figure}[t]
\includegraphics[scale=0.35]{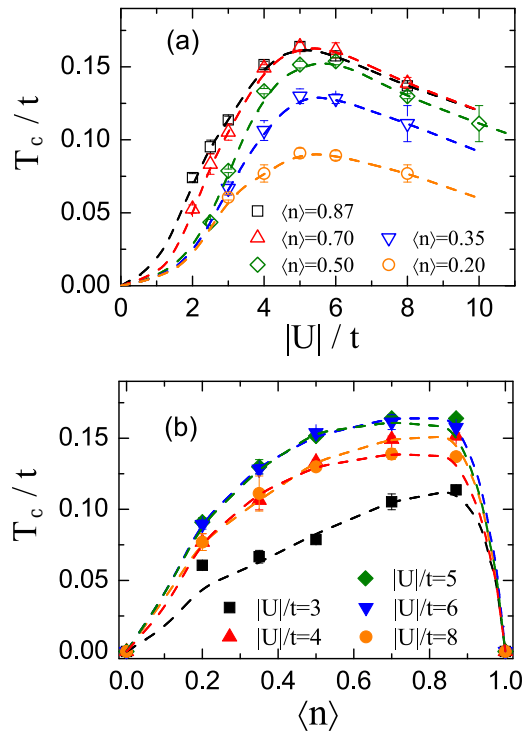} 
\caption{(Color online) Critical temperatures as function of (a) the interaction strength $|U|/t$, and (b) the electronic density $\langle n \rangle$. The curves are guides to the eye.}
\label{fig:Tc} 
\end{figure}

\begin{figure}[t]
\includegraphics[scale=0.28]{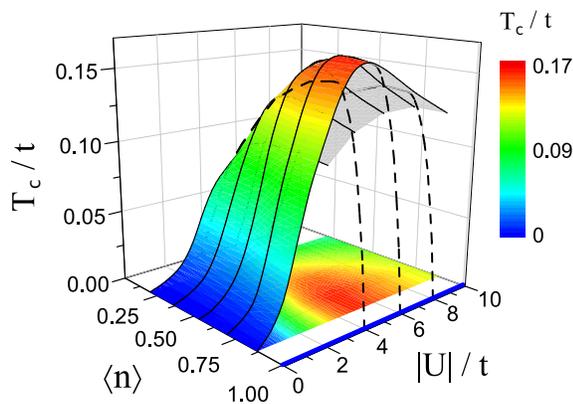} 
\caption{(Color online) Finite temperature phase diagram of the attractive Hubbard model in the square lattice.}
\label{fig:phase_diagram} 
\end{figure}

We map out the critical temperature for other values of $U$ and $\ave{n}$, making use of the weak dependence of the superfluid density with $L$: in what follows, most of the results for $T_c$ have been determined from simulations on lattices with linear size $L=14$ or 16.
Figure \ref{fig:Tc}\,(a) shows the critical temperature as a function of $U/t$, for different fermionic densities, and we note that it displays a maximum, $T^{\rm max}_{c}$, at some value, $U_m$,  which depends very weakly on $\ave{n}$ within the range considered here; we will return to this point below, in connection with the BCS-BEC crossover.
It is also instructive to examine the dependence of $T_{c}$ with the electronic density, for fixed values of $U$, with the results shown in Fig.\,\ref{fig:Tc}\,(b). 
We see that for each fixed $U$, $T_c$ displays a broad maximum for $0.7 \lesssim \ave{n}\lesssim 0.9$, and it drops sharply to zero at half filling by virtue of the Mermin-Wagner theorem; accordingly, in three dimensions $T_c(\ave{n})$ for fixed $U$ displays a maximum around $\ave{n}\approx 0.9$, but reaches a finite value at $\ave{n}=1$ \cite{dosSantos94}. 
We also provide the estimates for the critical temperature in tabular form (see Table \ref{tab:allT}), while the location of the `sweet spot' for $T_c$ is highlighted in Figure \ref{fig:phase_diagram}.  
  
\begin{figure}[t]
\includegraphics[scale=0.3]{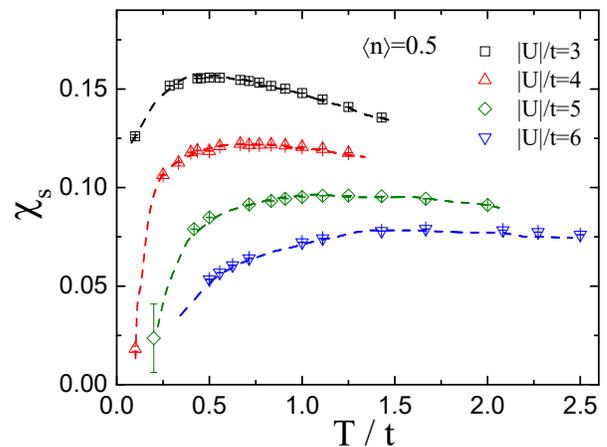} 
\centering
\caption{The uniform spin susceptibility as a function of temperature,  for different values of the on-site attraction, $|U|$, and at quarter filling, $\ave{n}=0.5$, for a linear lattice size $L=14$. The curves are guides to the eye.}
\label{fig:susc_n05} 
\end{figure}

\subsection{Pairing  temperature}
\label{ssec:Tp}

As mentioned in the Introduction, the pairing temperature provides a temperature scale around which Cooper pairs are formed; the pair-breaking gap is therefore expected to be related to a gap in spin  excitations, which, in turn, may be detected as a downturn in the uniform magnetic susceptibility, $\chi_s$, as the temperature is 
lowered~\cite{Randeria92,dosSantos94,Magierski09,Magierski11,Wlazlowski13,Tajima14}.  

Accordingly, Fig.\,\ref{fig:susc_n05} shows our DQMC data for the temperature dependence of the uniform susceptibility, $\chi_s$ [see Eq.\,\eqref{eq:magsusc}], at quarter filling and for different strengths of the attractive interaction. 
We first note that the magnitude of $\chi_{s}$ decreases with increasing $|U|$, following the trend predicted within RPA, $\chi^\text{RPA}=\chi_{0}/(1 + |U|\chi_{0})$.
In addition, for fixed $U$ we see that  $\chi_{s}$ drops steadily below some temperature, whose location depends on $U$. 
In line with the idea that this downturn in $\chi_s$ signals the formation of local pairs within some temperature scale~\cite{Randeria92,dosSantos94,Magierski09,Magierski11,Wlazlowski13,Tajima14}, we adopt the position of the maximum in $\chi_s$ as the pairing scale, $T_p(U)$; the fact that the maximum of $\chi_s$ can be quite broad is also consistent with the idea of a crossover, or a temperature scale, instead of a sharp transition. 

\begin{figure}[t]
\includegraphics[scale=0.31]{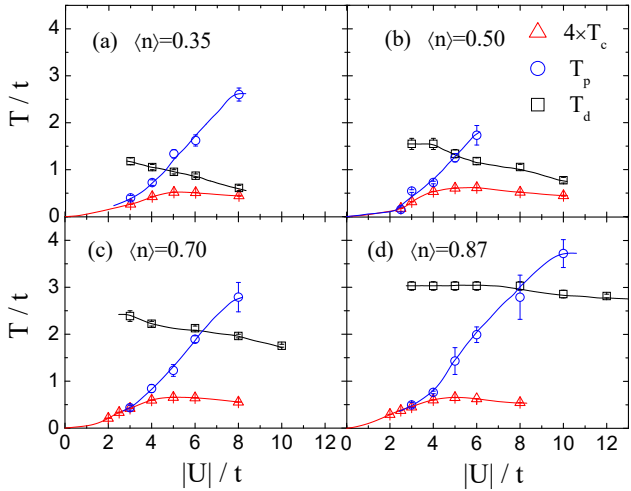}
\caption{Critical ($T_c$), pairing ($T_p$), and degeneracy ($T_d$) temperatures (in units of $t$) as functions of the  interaction strength $|U|/t$, obtained from DQMC simulations on a lattice with linear size $L=14$, and different band fillings. The curves are guides to the eye.
}
\label{fig:Tc_Tp_Td_difn} 
\end{figure}

By repeating this procedure for different band fillings, we generate the plots $T_p(U,\ave{n})$ shown in blue in Figs.\,\ref{fig:Tc_Tp_Td_difn}(a)-(d); these data are also displayed in Table \ref{tab:allT}.
For comparison of trends, we also include $T_c(U,\ave{n})$ in each panel of Figs.\,\ref{fig:Tc_Tp_Td_difn}(a)-(d).
We see that the difference between $T_p$ and $T_c$ gets smaller as $U$ decreases, which is a manifestation of the fact that in the BCS regime Cooper pairs are formed and condense at the same temperature. 
By contrast, for large $|U|$, pairs are formed at temperatures much higher than the condensation temperature: $T_p\sim |U|$, while $T_c\sim  |U|^{-1}$.

\subsection{Degeneracy temperature}
\label{ssec:Td}

\begin{figure}[t]
\includegraphics[scale=0.28]{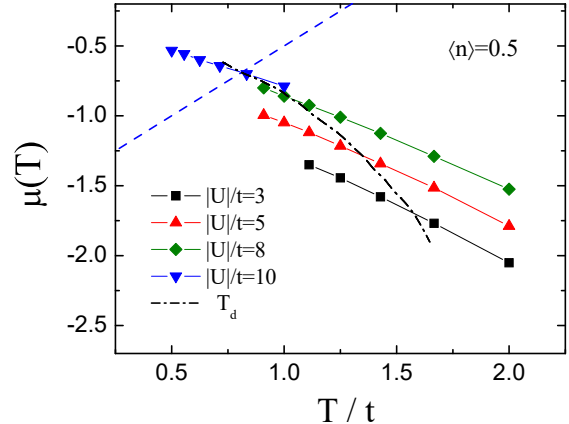} 
\caption{Each set of data points represents the temperature dependence of the chemical potential required to keep a constant fermionic density, $\ave{n}=0.5$, for a given $U$. 
The (blue) dashed line is the right-hand side of Eq.\,\eqref{eq:Td} for $|U|/t=10$, whose intersection with the $|U|/t=10$ data points determines $T_d$ for this particular $U$. 
The (black) dash-dotted line is the locus of the intersections for different values of $|U|$.
}
\label{fig:chemic_n05} 
\end{figure}

At high temperatures, the fugacity of a Fermi gas is small, $z\ll1$, while deep in the degenerate regime, fully dominated by quantum effects, one has $z\gg1$. 
We may therefore define a temperature scale for degeneracy, $T_{d}$, as the one in which $\ln z\sim 1$, i.e.~$ \mu = k_B T $. 
However, we note that in dealing with tight-binding fermions on a lattice, the bandwidth is finite and shifted from the continuum parabolic band. 
Furthermore,
due to the Hubbard term in the Hamiltonian, the Hartree shift must be taken into account when defining $T_{d}$~\cite{Randeria92}. 
Therefore, the degeneracy temperature is given by the solution of
\begin{equation}
	 \mu (T)  = k_B T - 4t - \frac{|U|}{2}(\ave{n} -1) ~,
	\label{eq:Td}
\end{equation}
for fixed $\ave{n}$ and $U$.
For instance, the data points in Fig.\,\ref{fig:chemic_n05} represent the temperature dependence of the chemical potential giving rise to $\ave{n}=0.5$, for different values of $U$, while the (blue) dashed line is the right-hand side of Eq.\,\eqref{eq:Td} for fixed $|U|/t=10$.  
Thus, the degeneracy temperature as a function of $|U|$ for a given $\ave{n}$ is obtained by extracting the points of intersection between the dashed curves (one for each value of $|U|$) and the corresponding $\mu(T)$ curves; the final outcome for this filling is displayed as the (black) dash-dotted line in Fig.\,\ref{fig:chemic_n05}.

The results for the degeneracy temperature appear in Figs.\,\ref{fig:Tc_Tp_Td_difn}(a)-(d), as well as in Table \ref{tab:allT}. 
For fixed $\ave{n}$, we see that $T_d$ decreases with $|U|$, while for fixed $|U|$ it increases with $\ave{n}$.
The relative positions between $T_d$ and $T_p$ in the different regimes of the pairing interaction allows us to form an intuitive picture of the mechanisms at play. 
First, we note that as the temperature is lowered in the weaker coupling part of the diagrams, fermionic particles first enter into a degenerate Fermi liquid regime, then they pair up, and finally condense into a superfluid at lower temperatures.
By contrast, in the strong coupling region the strength of the interaction forces fermions to first pair up forming bosonic particles before they enter into the degenerate regime at a lower temperature. 
In this regime, the effective density of unpaired fermions is smaller than the nominal $\ave{n}$, so that a smaller temperature is required to make their wave packets overlap. 
At a given temperature, the unpaired fermions act mostly as glues mediating the formation of the superfluid condensate (more on this below).

\section{Characterization of the BCS-BEC crossover}
\label{sec:Xover}

While there is consensus over the main qualitative differences between the BCS and BEC regimes, a quantitative characterization is still lacking, especially highlighting quantities accessible through quantum gas microscope measurements in optical lattices.   
Having established the different temperature scales, we now discuss some quantities which could be followed throughout the crossover.   

\begin{figure}[t]
\includegraphics[scale=0.4]{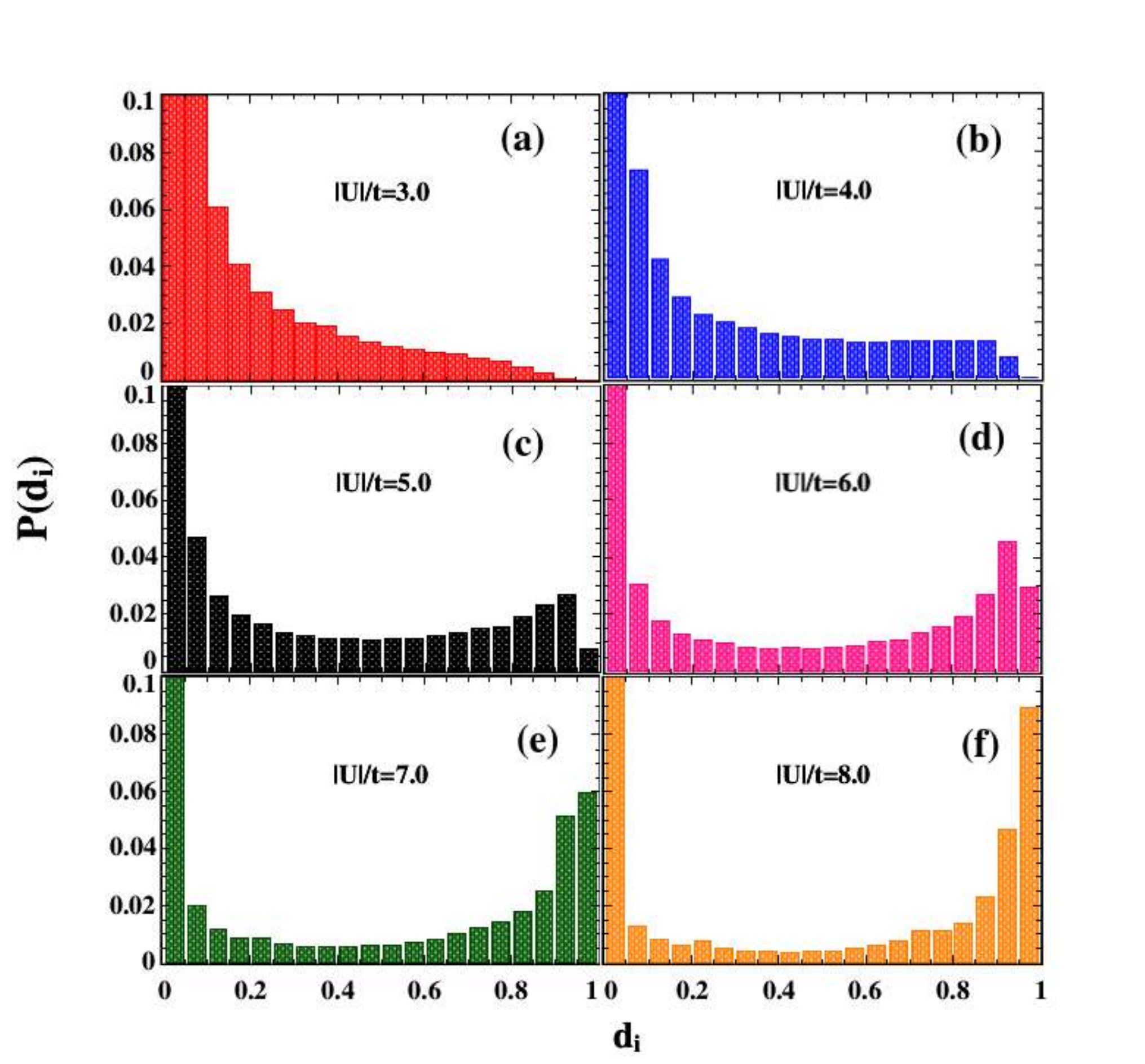} 
\caption{(Color online) Histograms of the normalized statistical weight of the double occupancy, for different values of the attractive interaction $U/t$, for fixed fermionic density, $\ave{n}=0.5$, temperature, $T/t = 0.2$, and lattice size, $L=14$. }
\label{fig:DO_05beta50}. 
\end{figure}

\begin{figure}[h]
\includegraphics[scale=0.4]{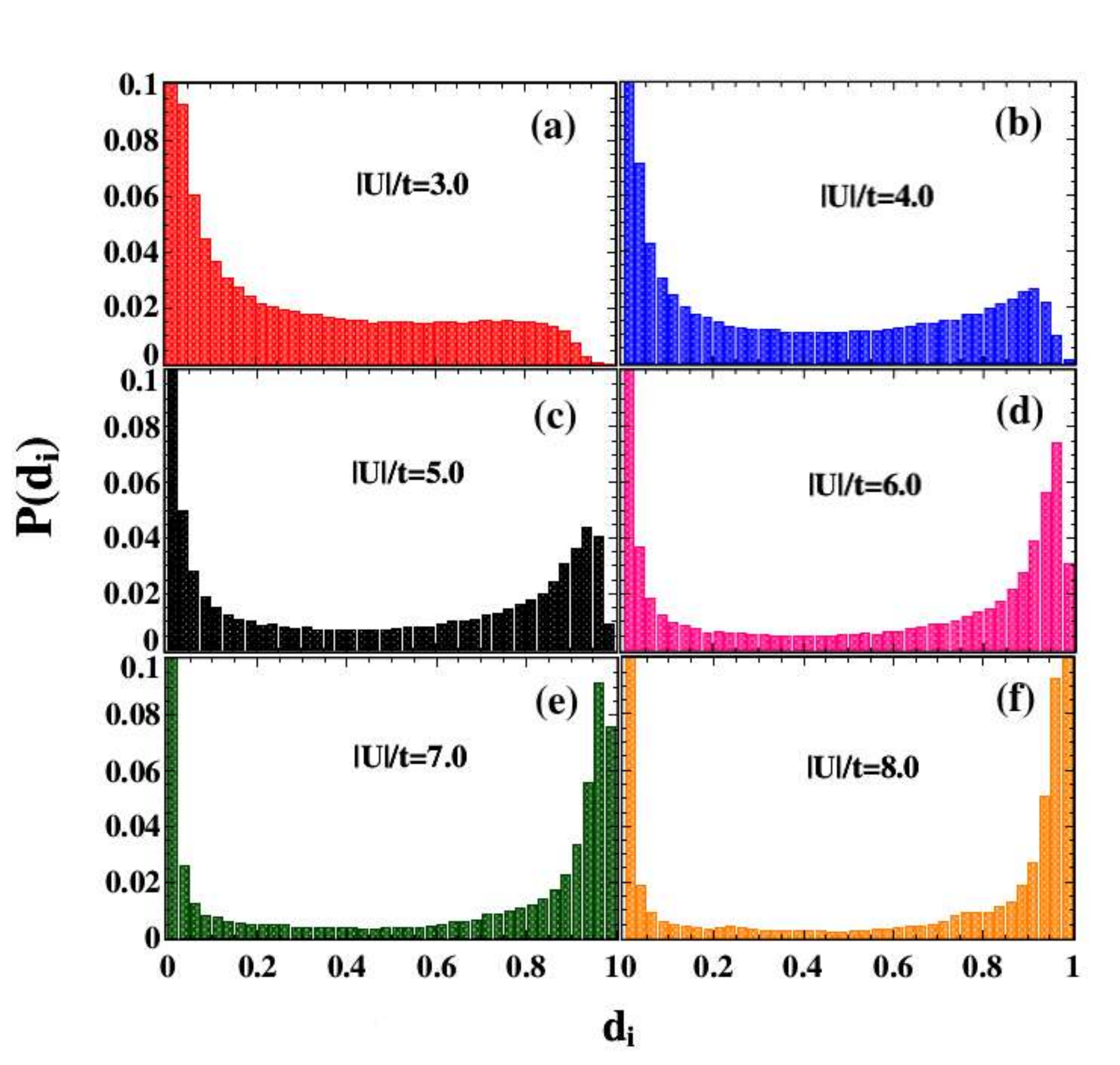} 
\caption{(Color online) Same as Fig.\,\ref{fig:DO_05beta50}, but for density, $\ave{n}=0.87$.
}
\label{fig:DO_087beta50} 
\end{figure}

The average double occupancy on a given site is defined as
\begin{equation}
	d_{\iv}=\ave{n_{\iv\uparrow}n_{\iv\downarrow}},
\label{eq:di}	
\end{equation}	
and ranges from 0 to 1. 
We note that in the extreme limit of $|U|\to\infty$ and $T\to0$, sites would be either doubly occupied by fermions or empty: a distribution of $d_\iv$ would be peaked at both $d_\iv=0$ and $d_\iv=1$.
In the opposite limit of weak coupling, $d_\iv$ should be strongly peaked at $d_\iv=0$.  
Accordingly, Figs.\,\ref{fig:DO_05beta50} and \ref{fig:DO_087beta50} follow the evolution of the double-occupancy distribution with the strength of attraction, at a fixed temperature, but for different band fillings. 
For $n=0.5$, a peak near $d_\iv=1$ starts developing at $|U|/t \approx 4.5 \pm 0.5$, and as $|U|$ increases this peak becomes more pronounced while moving towards $d_\iv=1$. 
For $n=0.87$, the $d_\iv=1$ peak starts developing at smaller values of $|U|$, namely at $|U|/t \approx 3.5 \pm 0.5$.
The reason for this decrease in $|U|$ must be attributed to the larger number of fermions available to pair up.
This also explains the fact that for a given $|U|$, it is more likely to find doubly occupied sites at larger fermionic densities.

\begin{figure}[t]
\centering
\includegraphics[scale=0.17]{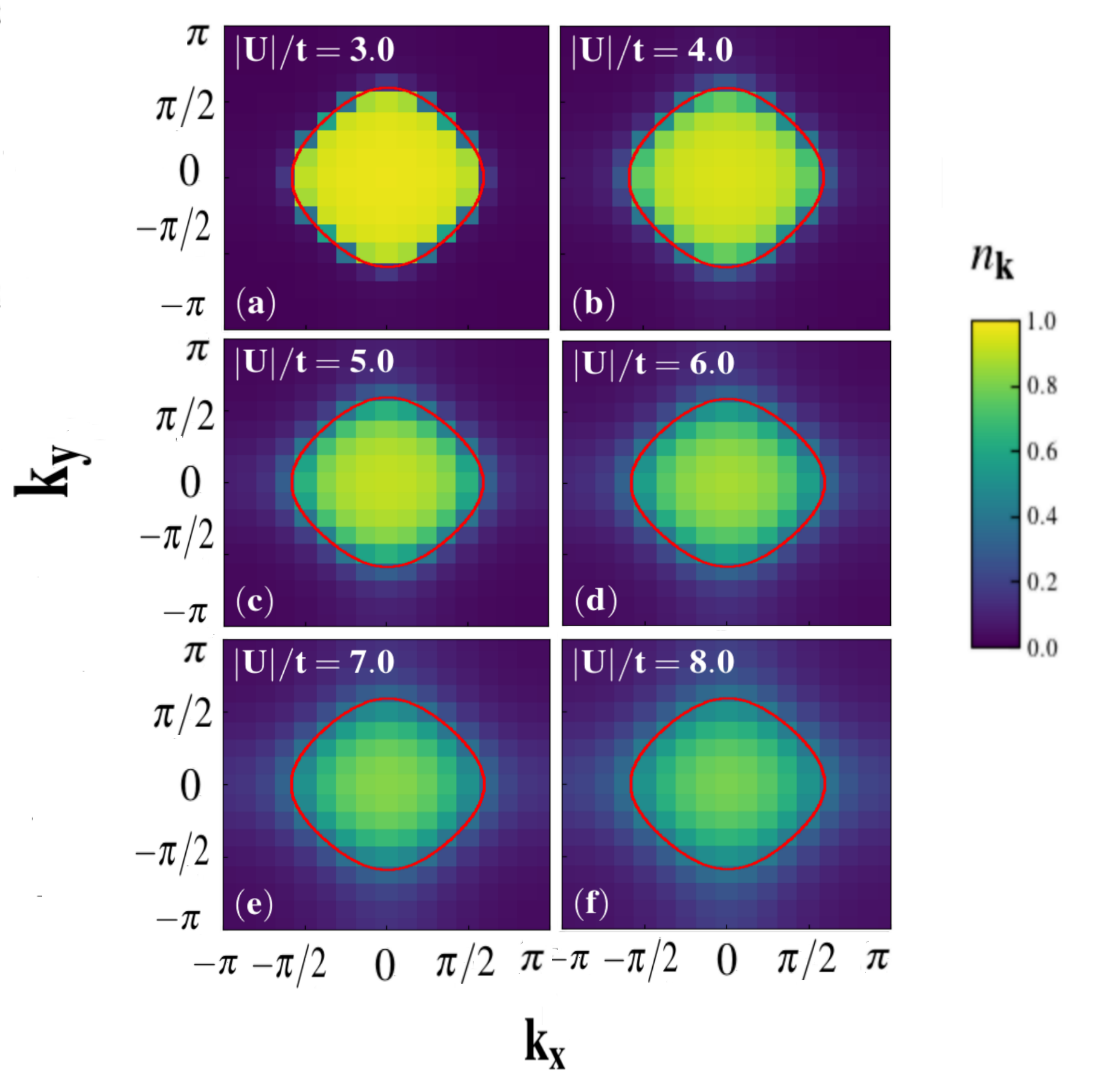}
\caption{Contour plot of momentum distribution for different values of $|U|/t$. 
Data are for density $\langle n \rangle=0.5$, temperature $T /t\approx 0.42$, and linear lattice size $L=16$. The red curve is the non-interacting Fermi surface for the same electronic density.
}
\label{fig:occup_Kspace} 
\end{figure}

Another interesting crossover probe is the density distribution in momentum space, 
\begin{equation}
	n_{\kv}\equiv \langle c_{\kv\sigma}^{\dagger} c_{\kv\sigma}^{\phantom{\dagger}} \rangle,
\label{eq:Nk}	
\end{equation}	
where, for brevity, the spin index was omitted in $n_{\kv}$, since $n_{\kv\uparrow}=n_{\kv\downarrow}$ in the absence of a symmetry-breaking magnetic field.
The results are displayed in Fig.\,\ref{fig:occup_Kspace} for $\ave{n}=0.5$ and for increasing values of $|U|/t$.
We see that while on the weak coupling side of the crossover the distribution bears some resemblance with one with a Fermi surface, on the strong coupling side the fermions are distributed way beyond the non-interacting Fermi surface.
One may therefore regard this as a crossover between a Fermi liquid (FL) at weak coupling and a non-Fermi liquid (NFL) at strong coupling. 
Interestingly, the NFL regime seems to appear when the occurrence of double occupancy is significant: this can be interpreted as indicating that unpaired fermions are also strongly tied to the tightly bound bosonic quasiparticles, which renders the FL framework inapplicable.    

This can be put in a more quantitative way, by calculating the quasi-particle weight, given in a form amenable to DQMC simulations \cite{Moreo90,Arsenault12,Chen12,Liu18,Xu19},  
\begin{align}
Z &= \left( 1 - \frac{\partial \Sigma ' (\omega)}{\partial \omega}\bigg|_{\omega \rightarrow 0} \right)^{-1}\nonumber\\
	&\approx \left( 1 - \frac{ \text{Im} \left[\Sigma (i\omega_{n})\right]}{\omega_{n}}\bigg|_{\omega_{n} \rightarrow 0} \right)^{-1},
\label{eq:quasi_weight}
\end{align}
where $\Sigma'$ is the real part of the self-energy, $\Sigma$, which can be directly calculated through the Green's functions \cite{Moreo90}; $\omega_{n}$ are the Matsubara frequencies.

\begin{figure}[t]
\centering
\includegraphics[scale=0.25]{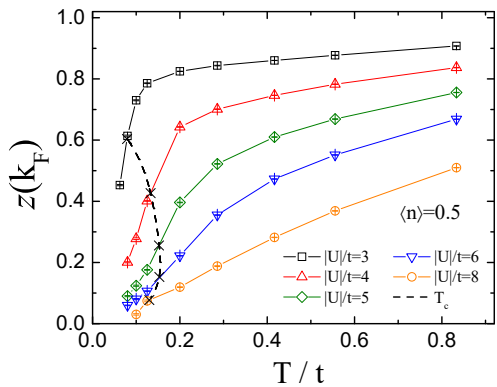}
\caption{Quasiparticle weight  as a function of temperature for different values of $|U|/t$ at quarter filling. The crosses define the values of $z(k_{F})$ at the respective critical temperatures, with the dashed line being a guide to the eye.
}
\label{fig:ZkF_with_Tc} 
\end{figure}

Figure \ref{fig:ZkF_with_Tc} shows the temperature dependence of the quasiparticle weight for $\ave{n}=0.5$. 
The locus of critical temperatures, $T_c(U)$, for the different values of $U/t$ reminds us that any attempt to identify a FL behavior for $T<T_c$ is doomed to failure, due to the opening of a superconducting gap. 
Nonetheless, we may draw some interesting conclusions from the behavior at $T > T_c(U)$: the decrease of the quasiparticle weight with decreasing temperature is steeper at strong coupling than at weak coupling, in line with the crossover FL-NFL alluded to in relation to Fig.\,\ref{fig:occup_Kspace}. 
Indeed, since the effective quasiparticle mass $m^* \simeq m/z(k_F)$, where $m$ is the bare fermionic mass, heavy quasiparticles are more strongly interacting, hence farther from the FL paradigm than light ones, or closer to NFL behavior.

\section{Conclusions}
\label{sec:conc}

Motivated by experimental attempts to investigate the superfluid transition of ultra-cold atoms in optical lattices with attractive on-site interactions, we have studied the region of optimal critical temperatures in terms of strength of interactions and fermionic density. 
By means of determinant quantum Monte Carlo simulations (DQMC), we have been able to pinpoint a somewhat wide region around $|U|/t \approx 5 \pm 1$ and $\ave{n} \approx 0.79 \pm 0.09$ with a critical temperature $T_c \approx 0.16t$, which, under the experimental conditions reported in Ref \cite{Mitra18}, amounts to $T_c \approx 8.8$ nK. 
We have also examined two other temperature scales, namely the degeneracy temperature and the pairing temperature. 
While the degeneracy temperature describes the region below which quantum effects dominate, the pairing temperature, $T_p$, sets the scale for the pair formation, which is believed to be closely related to the temperature for gapped spin excitations.
The fact that $T_p(U)$ does not show a strong dependence with $\ave{n}$ adds credence to its association with spectral properties.
We have also discussed possible scenarios for a breakdown of Fermi liquid (FL) theory across the BCS-BEC crossover, through analyses of DQMC data for the distribution of double occupancy, for the momentum distribution function, and for the quasiparticle weight.
The picture that emerges is that of a FL at weak coupling, which progressively breaks down when the dominant role played by unpaired fermions becomes that of mediating the interaction between tightly bound bosonic pairs.
The possibility of both BCS-BEC and FL-NFL crossovers taking place within the same range of interaction strengths, though appealing, cannot be ascertained at this point.

\section*{ACKNOWLEDGMENTS}

We are grateful to M.~Randeria for illuminating discussions and suggestions, as well as to J.P.~de Lima for his contributions at the initial stage of this work.
Financial support from the Brazilian Agencies CAPES, CNPq, FAPERJ and Instituto Nacional de Ciência e Tecnologia de Informa\c c\~ao Qu\^antica is also gratefully acknowledged.



\bibliography{Costa_attHub}
\end{document}